\documentclass[twocolumn, showpacs, longbibliography, prb,aps, amsmath, amssymb,superscriptaddress]{revtex4-1}
\usepackage{amsmath}  
\usepackage{amsfonts} 
\usepackage{graphicx} 
\usepackage[colorlinks,linkcolor=blue,citecolor=blue,urlcolor=blue]{hyperref}
\usepackage{ulem}
\usepackage{float}

\begin{document}

\title{{ Interaction-induced hopping phase in driven-dissipative coupled photonic microcavities}}

\author{S.R.K. Rodriguez} \email{said.rodriguez@lpn.cnrs.fr}
\affiliation {Laboratoire de Photonique et de Nanostructures (LPN), CNRS, Universit\'{e}
Paris-Saclay, route de Nozay, F-91460 Marcoussis, France}

\author      {A. Amo}
\affiliation {Laboratoire de Photonique et de Nanostructures (LPN), CNRS, Universit\'{e}
Paris-Saclay, route de Nozay, F-91460 Marcoussis, France}

\author      {I. Sagnes}
\affiliation {Laboratoire de Photonique et de Nanostructures (LPN), CNRS, Universit\'{e}
Paris-Saclay, route de Nozay, F-91460 Marcoussis, France}

\author      {L. Le Gratiet}
\affiliation {Laboratoire de Photonique et de Nanostructures (LPN), CNRS, Universit\'{e}
Paris-Saclay, route de Nozay, F-91460 Marcoussis, France}

\author      {E. Galopin}
\affiliation {Laboratoire de Photonique et de Nanostructures (LPN), CNRS, Universit\'{e}
Paris-Saclay, route de Nozay, F-91460 Marcoussis, France}

\author      {A. Lema\^{i}tre}
\affiliation {Laboratoire de Photonique et de Nanostructures (LPN), CNRS, Universit\'{e}
Paris-Saclay, route de Nozay, F-91460 Marcoussis, France}

\author      {J. Bloch}
\affiliation {Laboratoire de Photonique et de Nanostructures (LPN), CNRS, Universit\'{e}
Paris-Saclay, route de Nozay, F-91460 Marcoussis, France}
\affiliation {Physics Department, Ecole Polytechnique, F-91128 Palaiseau Cedex, France }

\date{\today}
\maketitle

\textbf{Bosons hopping across sites and interacting on-site are the essence of the Bose-Hubbard model (BHM)~\cite{Fisher}. Inspired by the success of BHM simulators with atoms in optical lattices,~\cite{Bloch12} proposals for implementing the BHM with  photons in coupled nonlinear cavities have emerged~\cite{Greentree06, Hartmann06, Angelakis, Carusotto09, LeBoite}. Two coupled semiconductor microcavities constitute a model system where the hopping, interaction, and decay of exciton polaritons --- mixed light-matter quasiparticles --- can be engineered in combination with site-selective coherent driving to implement the driven-dissipative two-site optical BHM. Here we  explore the interplay of interference and nonlinearity in this system, in a regime where three distinct density profiles can be observed under identical driving conditions. We demonstrate how the phase acquired by polaritons hopping between cavities can be controlled through effective polariton-polariton interactions. Our results open new perspectives for synthesizing density-dependent gauge fields~\cite{Carusotto11, Hafezi11, Fang12} for polaritons in two-dimensional multicavity systems.}\\

Understanding the emergence of collective phenomena in condensed matter systems is an example of a problem that quantum simulators may address. Ultracold atoms in optical lattices have enabled great progress in this direction~\cite{Bloch12}. Recently, photonic systems have been proposed for simulating the hopping and interaction of bosonic particles as described by the BHM, but in non-equilibrium conditions~\cite{Greentree06, Hartmann06, Angelakis, Carusotto09,  LeBoite}. In particular, driven-dissipative lattices of coupled nonlinear cavities can display strongly correlated steady-state phases characterized by the number of available stable modes~\cite{LeBoite}. For the minimal Bose-Hubbard system comprising two sites, i.e. a dimer,   intriguing quantum interference effects and single photon emission have been predicted\cite{Gerace09, Liew10}.  As we will show, the driven-dissipative Bose-Hubbard dimer (BHD)  displays striking phenomena  even at the mean-field level due to the interplay of interference, nonlinearity, and site-selective coherent driving.

Under time-harmonic driving of one site, the mean fields $\psi_{j}$ of the driven-dissipative BHD  are described by the coupled equations:
\begin{equation}~\label{eq1}
\begin{aligned}
 i \hbar \dot{\psi_1} &=  (\hbar \omega_1 - i\frac{\gamma_1}{2})\psi_1 + U|\psi_1|^2 \psi_1 - J\psi_2 + F e^{-i\omega t},   \\
 i \hbar \dot{\psi_2} &=  (\hbar \omega_2 - i\frac{\gamma_2}{2})\psi_2 + U|\psi_2|^2 \psi_2 - J\psi_1.
\end{aligned}
\end{equation}

\noindent $\omega_{j}$ and $\gamma_{j}/2$ ($j=1,2$)  are the on-site energy and decay rate, $J$ is the hopping energy, and $U$ is the interaction energy.  $F$ and $\omega$  are the driving amplitude and frequency on site $j=1$. The BHD dynamics  without driving ($F=0$) has been thoroughly studied with atoms, especially in relation to the self-trapping occurring when the total interaction energy, $U(N_1+N_2)$ with $N_j = |\psi_j|^2$ the mode populations, exceeds $J$~\cite{Smerzi97, Albiez05}. For dissipative (e.g. photonic) systems, the non-Hermiticity of the BHD  Hamiltonian~\cite{Graefe10} gives rise to distinct nonlinear phenomena. A dissipation-limited self-trapping time~\cite{Abbarchi13}, a dissipation-induced classical to quantum transition,~\cite{Raftery14} and spontaneous symmetry breaking~\cite{Hamel15}, have been observed with photons. With coherent driving on one site($F \neq 0$),  parametric instabilities ~\cite{Sarchi08} and nonclassical correlations~\cite{Gerace09, Liew10} have been predicted as hopping, interactions, and decay compete in setting a stationary state.  Despite impressive theoretical efforts in this direction, the driven-dissipative BHD has remained experimentally unreported with photons so far.

An excellent system for implementing the driven-dissipative  optical BHD comprises exciton polaritons in coupled semiconductor microcavities. Polaritons are hybrid light-matter quasiparticles formed by strong coupling between cavity photons and quantum well excitons~\cite{Weisbuch92}.  Polaritons can be confined and coupled by micro-patterning planar cavities, thereby acting on the photonic part of their wavefunction~\cite{Bayer98}. In this way, Hamiltonians describing molecular orbitals~\cite{Sala} or particles in lattices~\cite{Jacqmin} can be implemented. In addition, Kerr nonlinearities associated with the excitonic part of polaritons~\cite{Ciuti98} yield effective polariton-polariton interactions.  Steady-state nonlinearities such as bistability~\cite{Baas04} and polarization multistability~\cite{Paraiso10} have been observed in single cavities. Accessing the physics of the driven-dissipative non-equilibrium BHM requires  spatial coupling of nonlinear cavities which, in contrast to the coupling of the two polariton spin components~\cite{Paraiso10}, can include many degrees of freedom.  Here we take a first step in this direction by exploring a highly nonlinear polariton BHD.  We discover an interaction-induced phase for polaritons hopping between cavities. This  mechanism could enable the realization of non-Hermitian Hamiltonians with density-dependent gauge fields if extended to two-dimensional cavity arrays.

\begin{figure} \centerline{{\includegraphics[width=1\linewidth]{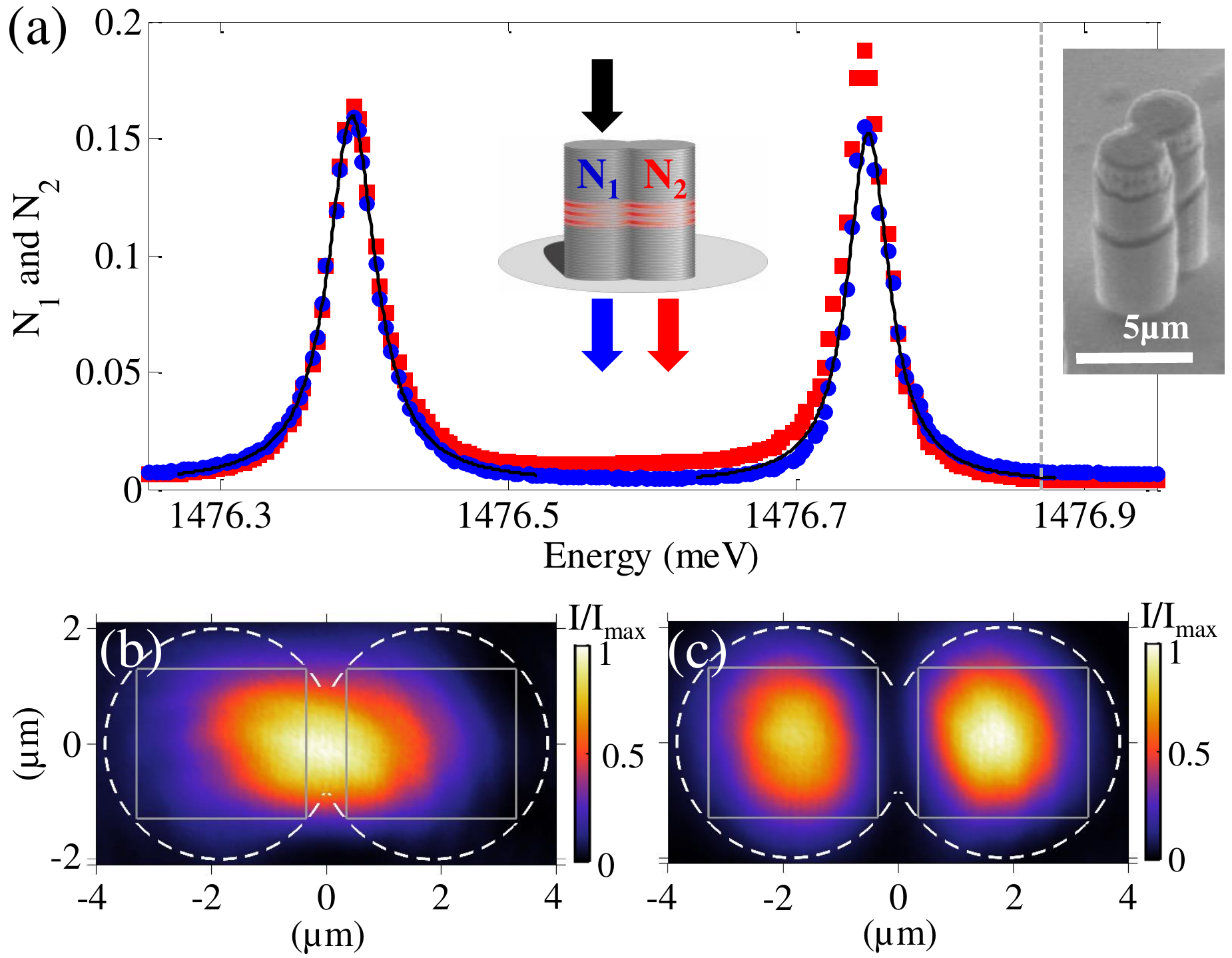}}}\caption{\textbf{Linear spectrum and eigenmodes.} (a) Mean polariton number in two coupled microcavities under continuous driving of one cavity.  The central inset illustrates the spatial and color code used throughout the manuscript: blue data points for the driven cavity on the left, and red data points for the undriven cavity on the right.  Black curves are Lorentzian fits. The dashed gray line indicates the driving energy used in Fig.~\ref{fig3} and  Fig.~\ref{fig4}. The right inset shows a scanning electron micrograph of the structure. (b) and (c) show the spatial distribution of the normalized transmitted intensity  at the low and high energy peaks corresponding to the bonding and antibonding resonances, respectively. The dashed lines delimit the two coupled microcavities. The solid lines indicate the integration area used to evaluate the population in each cavity throughout the manuscript. }\label{fig1}
\end{figure}

\begin{figure} \centerline{{\includegraphics[width=1\linewidth]{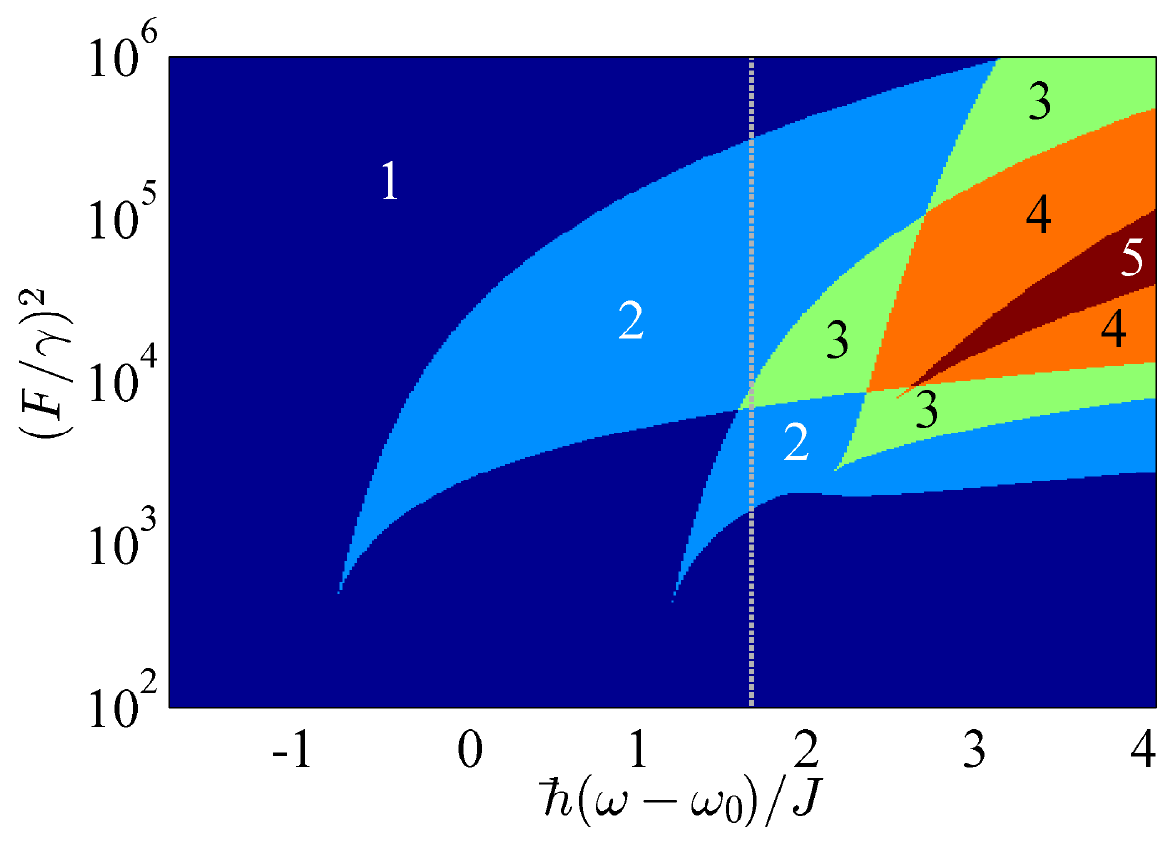}}}\caption{\textbf{Calculated number of stable modes.} The number of stable modes admitted by Equations~\ref{eq1} is shown in color as a function of the energy detuning $\hbar(\omega-\omega_{0})$ divided by the hopping energy $J$, and of the driving power ($F/\gamma$)$^2$ with $F$ the driving amplitude and $\gamma$ the loss rate of each cavity. Identical cavities with eigenfrequency $\omega_0$ are assumed. The dashed line indicates the driving energy used in Fig.~\ref{fig3} and  Fig.~\ref{fig4}}.\label{fig2}
\end{figure}

The two slightly overlapping cavities we investigate are shown in the Fig.~\ref{fig1}(a) inset.  The coupled cavities behave as a photonic molecule (PM), where strong coupling between polaritons in each cavity forms hybridized states~\cite{Bayer98}.  Figure~\ref{fig1}(a)  shows the linear spectrum of the PM. We drive the left cavity with a laser of variable frequency and quantify the cavity populations from spatially resolved transmission measurements (see Methods). The low and high energy peaks are the bonding and antibonding resonances of the PM, respectively. From Lorentzian fits to the spectra (black lines) we extract a bonding-antibonding splitting of $2J = 358 \pm 1$ $\mu$eV, well above the sum of the linewidths  $\gamma_B + \gamma_{AB} = 75 \pm 3$ $\mu$eV. Figure~\ref{fig1}(b) shows the  bonding mode, with  nonzero density at the center of the dimer reflecting the even parity of the wave function. In contrast, the antibonding mode in Fig.~\ref{fig1}(c) shows suppressed density at the center due to the odd  parity of the wave function.

To illustrate the wealth of nonlinear phenomena expected in the PM according to equations~\ref{eq1}, we present in Fig.~\ref{fig2} the calculated number of stable modes as a function of the dimensionless frequency detuning $(\omega-\omega_{0})/J$ and driving power $(F/\gamma$)$^2$.  We consider two identical cavities ($\omega_1 = \omega_2 =\omega_0$ and $\gamma_1 = \gamma_2 =\gamma$) with repulsive interactions ($U>0$) within each cavity. Figure~\ref{fig2} shows that for weak driving (negligible interactions here achieved for $(F/\gamma$)$^2 \lesssim 300$) and any  $\omega$,  or any power and $\omega-\omega_{0} < -0.8J$, the PM is monostable: there is a single input-output relation. When $-0.8J \lesssim \omega-\omega_{0} \lesssim 1.6J$,  the PM supports two stable modes, i.e., bistability.~\cite{Baas04} The monostable and bistable regimes are well-known;  their observation in our system is provided in the supplementary information. The third and most interesting driving condition is when $\omega-\omega_{0} \gtrsim 1.6J$, where  up to five stable modes can be observed. In the following, we show experiments and calculations for $\omega-\omega_{0} = 1.67J$ (dashed line in Fig.~\ref{fig2}), where one of the three accessible modes displays an interaction-induced hopping phase.

\begin{figure*} \centerline{{\includegraphics[width=\linewidth]{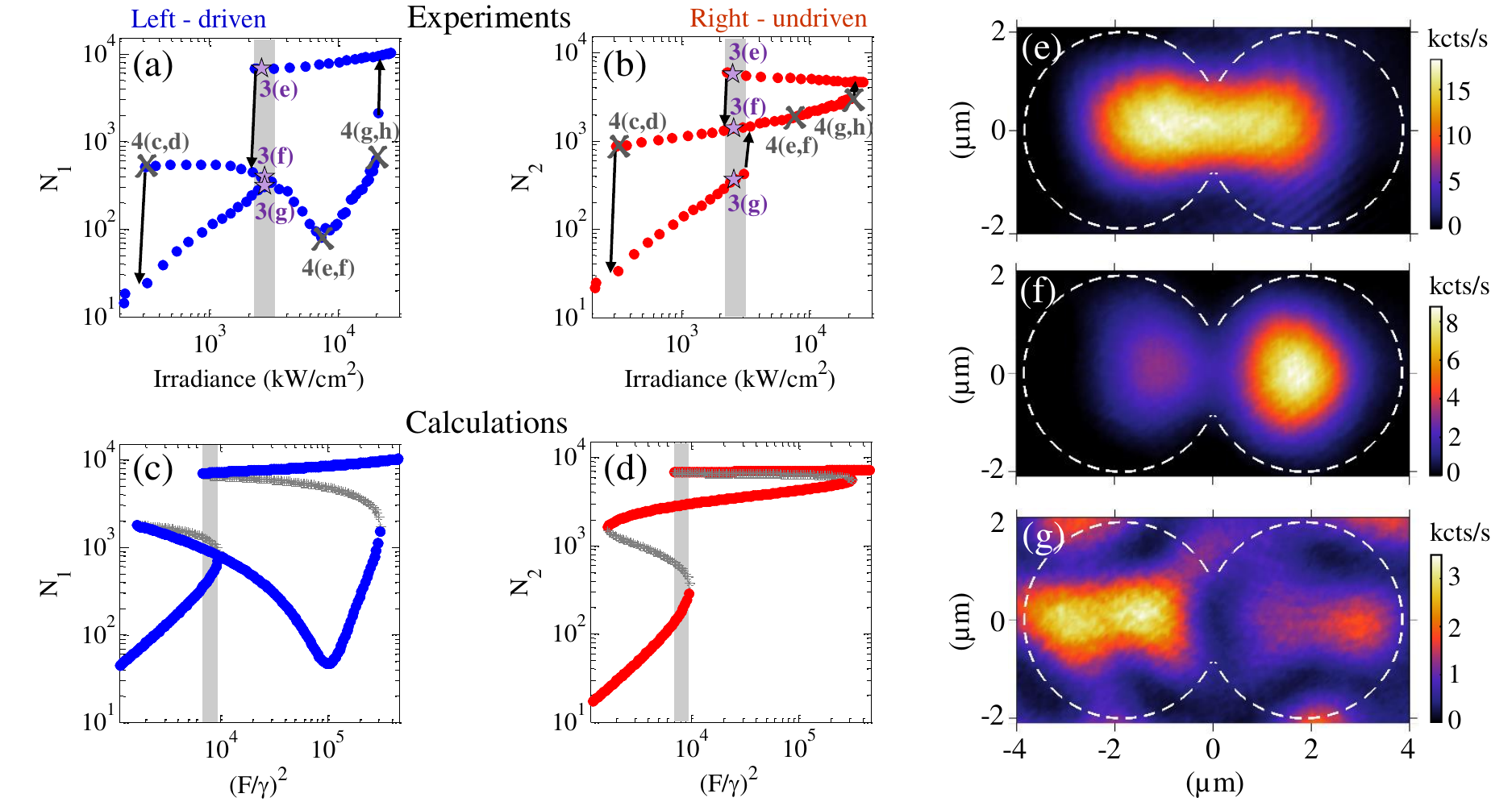}}}\caption{\textbf{Tristability.} (a) and (b) show the measured populations in the driven and undriven cavities, respectively, for 1476.87 meV driving energy [dashed line in Fig.~\ref{fig1}(a)].  Shaded areas indicate the irradiance range for tristability; (e,f,g) show three mode profiles obtained for the same irradiance as indicated with purple stars in (a) and (b). The gray crosses corresponds to the irradiance corresponding to Figs.~\ref{fig4}(b,c);  (c) and (d) show the populations calculated with the model described in the text. Blue and red lines correspond to the stable modes of the driven and undriven cavities, respectively, and gray lines correspond to unstable modes.}
\label{fig3}
\end{figure*}

Figures~\ref{fig3}(a,b) show experiments where the left cavity is driven at an energy of 1476.87 meV [dashed line in Fig.~\ref{fig1}(a)]. We observe a  pronounced hysteresis in the populations as a function of the irradiance. The hysteresis involves three branches. Changes in density are reversible along each branch.  In the shaded region in Figs.~\ref{fig3}(a,b), the PM is tristable: three different  stable density profiles can be observed at the same irradiance. Figures~\ref{fig3}(e),~\ref{fig3}(f), and ~\ref{fig3}(g) illustrate three density profiles at the same irradiance, indicated by the stars in Figs.~\ref{fig3}(a,b). Which one of these profiles is observed depends on the history of the system, or in which direction the irradiance is scanned.

The nonlinear jumps and the branches in Figs.~\ref{fig3}(a,b) can be understood by comparing the total interaction energy $U(N_1 + N_2)$ with the energy detuning between the laser and the linear eigenmodes of the system.  At the first upwards threshold, the  antibonding mode blueshift brings it in resonance with the laser.  The signature of the antibonding mode --- a suppressed population at the center of the dimer --- can be recognized throughout the middle branches, as illustrated in  Figure~\ref{fig3}(f).   For greater irradiance the second upwards threshold brings the bonding mode in resonance with the laser. This sets the populations into the highest branches, where the features of the bonding mode can be recognized [see the mode profile in Fig.~\ref{fig3}(e) resembling  the linear bonding mode in  Fig.~\ref{fig1}(b)]  We stress that these are all qualitative similarities, since bonding and antibonding  are linear eigenmodes of the system. In the supplemental information we show how the spectrum of the PM evolves in the nonlinear regime.

The measurements in Figs.~\ref{fig3}(a,b) are qualitatively reproduced using equations~\ref{eq1}. Figures~\ref{fig3}(c,d) show calculated populations using parameters deduced from the fits to the linear spectrum (see Methods). Besides the three stable branches observed in experiments, the calculations show two unstable branches (gray lines) not accessed in experiments. These unstable branches emerge when a fixed point loses its stability and a new fixed point is created~\cite{Sarchi08}. Beyond the qualitative agreement, the calculations show some differences with experiments. These are likely due to power fluctuations in the driving laser, which make it difficult to access the end-points of the branches where instabilities take place. Further differences stem from the fact that in theory,  i) the populations in the driven and undriven cavities are perfectly separable, and ii) the driving force acts on one cavity only. Both i) and ii) are not strictly true in experiments due to the spatial overlap of the cavities and the finite beam waist.

\begin{figure*} \centerline{{\includegraphics[width=0.95\linewidth]{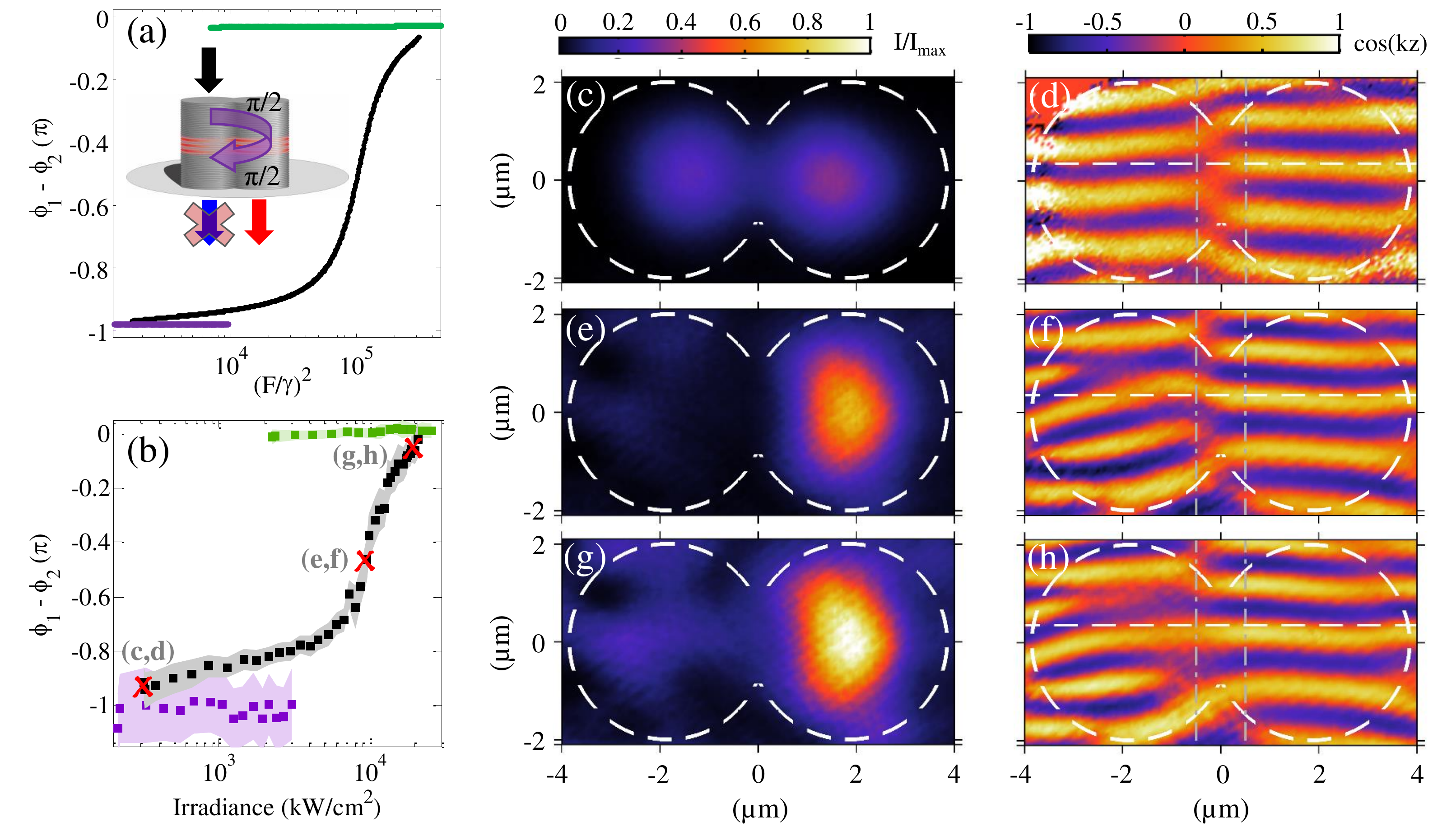}}}\caption{ \textbf{Interaction-induced hopping phase.}  (a) Calculated, and (b) measured,  phase difference between the intracavity fields for the three stable branches in Fig~\ref{fig3}.  The phase difference $\phi_1 - \phi_2$  is shown in purple for the low density branches, in green for the high density branches, and in black for the middle branches (where driven cavity exhibits a dip in the population). The shaded areas in (b) of the same color as the data points correspond to $2\sigma$ ($\approx 95\%$) confidence intervals on the fits performed to retrieve the phase of each cavity (see Methods). (c)-(h) show three representative density (central panels) and normalized interferogram (right panels) plots along the middle branch. The effective hopping phase $\phi_1 - \phi_2$ and irradiance corresponding to each one of these three cases is marked by a red cross in (b).  The color bar on top of panel (c) applies to all three density plots (c,e,g), while the color bar on top of panel (d) applies to all three normalized interferograms in (d,f,h). The vertical dash-dotted lines in (d,f,h) indicate where cosine functions were fitted to the normalized interferogram to retrieve the phase of each cavity. The horizonal dashed lines in (d,f,h) are all at the same position and serve as guides to the eye. }\label{fig4}
\end{figure*}

A striking feature in Figs.~\ref{fig3}(a,c) is the pronounced population dip along the middle branch. The occurrence of this dip between the two upwards thresholds and its absence in the undriven cavity suggests that this is an interference effect. To elucidate the underlying mechanism, we calculate in Fig.~\ref{fig4}(a) the difference $\phi_1-\phi_2$ between the phases of the field in each cavity (see Methods). This is the phase picked up by a polariton hopping between cavities. Since polaritons must hop  twice to interfere with the driving field in the first cavity, the stationary population depends on the round-trip phase $2(\phi_1-\phi_2)$.  Figure~\ref{fig4}(a) shows that  $2(\phi_1-\phi_2)\approx0$ (modulo $2\pi$) for the lowest and upper branches, irrespective of the driving strength. These  branches correspond to the lowest and highest branches in Figs.~\ref{fig3}(a,b), where interference in the driven cavity is constructive.  Notice that for the lowest (resp. upper) branch, $\phi_1-\phi_2\approx -\pi$ (resp. $0$), which is the characteristic phase relation of the antibonding (resp. bonding) mode.  Interestingly, for the middle branch in Fig.~\ref{fig4}(a), $\phi_1-\phi_2$ varies from $-\pi$ to $0$.  Therefore, the round-trip phase makes the interference in the driven cavity  change from constructive to destructive and back to constructive for increasing intensity.

We performed power-dependent interferometry measurements to directly observe the predicted interaction-induced hopping phase. For this purpose, the cavity transmission was interfered with an expanded section of the excitation laser beam (see Methods). Next,  we fitted cosine functions to the normalized interferogram in each cavity. Figure~\ref{fig4}(b) shows the difference between the fitted phases, $\phi_1-\phi_2$, in good agreement with our calculations.  Figures~\ref{fig4}(c)-~\ref{fig4}(h) show representative density (left panels) and interferogram (right panels) plots along the middle branch [black squares in Fig.~\ref{fig4}(b)].   Figures~\ref{fig4}(c,d) and ~\ref{fig4}(g,h) show a significant density in the driven cavity  when $\phi_1-\phi_2\approx -\pi$ and $\phi_1-\phi_2\approx 0$, respectively; these are conditions of constructive interference.  In contrast, Figs.~\ref{fig4}(c,d) shows that the driven cavity is dark at the destructive interference condition $\phi_1-\phi_2=-\pi/2$, i.e. a round trip phase of $-\pi$. The observation of this density-dependent interference demonstrates that the hopping phase can be optically controlled through interactions.

Beyond the BHD, an interaction-controlled hopping phase in two-dimensional lattices could enable the exploration of BHMs with density-dependent gauge fields. The proposed extension  relates to the seminal work by Aharanov \& Bohm~\cite{AB59}, and Berry~\cite{Berry84},  who realized that a nonzero phase acquired by  a particle in a closed-loop trajectory implies the existence of a nonzero vector potential $\mathbf{A}$. Specifically, the phase acquired when hopping from site $i$ to $j$  can be expressed as $\phi_{i,j} = \frac{e}{h} \int_{\mathbf{r}_i}^{\mathbf{r}_j} \mathbf{A} \cdot d \mathbf{l}$, where $e$ is the elementary charge~\cite{Dalibard11}.  Thus, synthethic magnetism~\cite{Carusotto11, Fang12} and topologically non-trivial states~\cite{Hafezi11, Rechtsman13} could be achieved for photons in two-dimensional arrays of coupled nonlinear cavities by engineering an interaction-induced hopping phase.\\

\noindent {\large \textbf{Methods}}

\noindent\textbf{Sample.}
The planar cavity was grown by molecular beam epitaxy and comprises a $\lambda/2$ GaAs cavity between two Ga$_{0.9}$Al$_{0.1}$As/Ga$_{0.05}$Al$_{0.95}$As distributed Bragg reflectors with 26 and 30 pairs for the top and bottom one, respectively.   One 80 \r{A}-wide InGaAs quantum well with an exciton energy of 1480.7 meV  is positioned at the center of the cavity.   Strong exciton-photon coupling leads to a Rabi splitting of  $3.4$ meV. The coupled microcavities are fabricated by electron beam lithography and dry etching of the planar cavity. Based on the linear transmission spectra,  we estimate a polariton ground state energy for each microcavity of $1476.6$ meV and a linewidth of  $37.5\pm1.5 \mu$eV. Based on the polariton dispersion in the planar cavity  (see supplemental information), we estimate a photon fraction $|C|^2$ = $0.84 \pm 0.03 $ ($C$ being the photonic Hopfield coefficient) at the driving energy of the experiments in Fig.~\ref{fig3} and Fig.~\ref{fig4}.

\noindent \textbf{Experiment.}
All experiments are performed at 4 K in transmission geometry, collecting the driving laser transmitted intensity from the substrate side. The  laser is a tunable MSquare Ti:Sapphire oscillator with $<10$ MHz  linewidth.  The excitation and collection objectives have a numerical aperture of 0.5 and 0.4, respectively. The excitation laser beam is linearly polarized parallel to the dimer axis [horizontal line at $0$ $\mu$m in all density and interference color plots].

We quantify the population in each cavity as follows. The transmitted intensity is recorded with a CCD camera without any spatial or spectral filtering. The counts detected within the left and right squares delimited by the gray solid lines in Fig.~\ref{fig1} are attributed to the driven and undriven cavities, respectively. The count rate for each cavity $n_j$ is converted to the polariton population $N_j$ via the following relation: $N_j= 2 n_j \tau \Phi^{-1} |C|^2$. The factor of 2 takes into account that roughly half of the population decays in the direction opposite to the detector, $\tau=18$ ps is the polariton lifetime, $\Phi$ is the detection efficiency (including collection),  and $|C|^2$  quantifies the fraction of polaritons that decay radiatively.

For the measurements in Fig.~\ref{fig4} we used a Mach-Zender interferometer as described next. The first beam splitter directed half of the power in the driving laser to the coupled cavities, and  the other half of the power by-passed the cavities and served as a reference. The reference beam was expanded, making its beam waist at the position of the detector about three times the diameter of the cavity. The intensity in the reference beam was controlled with a neutral density filter. We recorded the transmitted intensity by the coupled cavities $I_{c}$, and the intensity in the reference beam $I_{r}$.  Next, $I_{c}$ and $I_{r}$ were combined  by a second beam splitter placed between the output of the cavities and the detector. We call the combined total intensity $I_t$. The data was analyzed with the two-beam interference equation  $I_t = I_c + I_r + 2\sqrt{I_c I_r}$ cos($kz$). The quantity $kz$ corresponds to the optical path difference between the two arms of the interferometer, which is controlled by the position of the second beam splitter and the alignment of the two beams. Figures~\ref{fig4}(d,f,h) plot the cos($kz$) term in color as a function of space. We call this quantity, bounded between -1 and 1, the normalized interferogram. To retrieve the interaction-induced hopping phase in Fig.~\ref{fig4}(b), we repeated this procedure while scanning the driving power along all three branches. Next, we analyzed the normalized interferogram as follows. We took cuts of the interferogram along vertical lines [dash-dotted lines in Figs.~\ref{fig4}(d,f,h)] at a  distance of $\pm 0.5$ $\mu$m from the center of the dimer. The cuts at $- 0.5$ $\mu$m  correspond to the driven cavity, and the cuts at $+ 0.5$ $\mu$m correspond to the undriven cavity.  To each cut we fitted a function of the form $A_j$cos($B_j y + \phi_j$) + $C_j$, where $A_j$, $B_j$, $\phi_j$, and $C_j$ are fit parameters corresponding to the $j^{th}$ cavity  ($j=1,2$), and $y$ is the vertical dimension. Figure~\ref{fig4}(b) plots the difference between the fitted phases $\phi_1 - \phi_2$.  The behavior reported for  $\phi_1 - \phi_2$  at $\pm 0.5$ $\mu$m is robust over distances greater than 1  $\mu$m  with respect to the center of the dimer. For larger distances, the phase patterns in Figs.~\ref{fig4}(f,h) exhibit dislocations where the mode intensity vanishes near the walls of the driven cavity.  The origin of these dislocations is the presence of parasitic scattered laser light which interferes with the weak cavity transmission  at the detector.  Due to the highly nonlinear transmission through the driven cavity, the contribution of the parasitic light can be more than 2 orders of magnitude greater  at high irradiance than at low irradiance.  The contribution of the parasitic light to the measured phase patterns is only significant in the regions of vanishing mode intensity at the edges of the driven cavity. Hence, the measured phase patterns in these regions do not reflect the intracavity field phase only.  For these reasons we consistently analyze the phase patterns in Fig.~\ref{fig4} far from these artefacts and near the center of the dimer, i.e., at $\pm 0.5$ $\mu$m.\\

\noindent \textbf{Calculations.}
For all calculations we seek the stationary solutions $\psi^s_j$  ($j=1,2$) to the differential equations~\ref{eq1}. We start by inserting the ansatz $\psi(t) = \psi^s_j e^{-i \omega t}$ in equations~\ref{eq1}. This leads to the algebraic equations

\begin{equation}
\begin{aligned}
(\hbar \omega_j -\hbar \omega - i\frac{\gamma_j}{2})\psi_j^s + U N_j \psi_j^s - J\psi_{3-j}^s + \delta_{j1}F = 0,
\end{aligned}
\end{equation}

\noindent where $N_j = |\psi_j|^2$ are the mode populations.  The populations are obtained by writing the above equations as a polynomial in powers of $N_2$, calculating the roots of that polynomial, and then inserting the solutions in the remaining equation to obtain $N_1$.   The phase difference between the intracavity fields, $\phi_1 - \phi_2$, is calculated  by inserting the populations in the supplementary equations $\psi^s_j = \sqrt{N_j} e^{-i \phi_j}$.  Finally, we assess the stability of the stationary solutions by analyzing the spectrum of small fluctuations in their vicinity, i.e.  $\psi_{j}(t) = [\psi_j^s + \delta \psi_j(t)] e^{-i \omega t}$.  This is performed following the procedure outlined by Sarchi \emph{et al.}\cite{Sarchi08}. Sections 2 and 3 from Ref.~\onlinecite{Sarchi08} describe the physics of the model we employ throughout this manuscript (equations~\ref{eq1}), including the various kinds of stable solutions and instabilities that the two coupled nonlinear modes support. However, the analysis therein is restricted to the populations of the two modes and not to their relative phases. The phase analysis in Fig.~\ref{fig4}(a) and the counting of the number of modes in Fig.\ref{fig2} are results from the present work.

Based on Lorentzian fits to the measured linear spectrum, we set $\hbar \omega_1 = \hbar \omega_2 = 1476.6$ meV, $\gamma_1 = \gamma_2 = \gamma = 37.5$ $\mu$eV, and $J=179$ $\mu$eV for all calculations.  $U=0.07$ $\mu$eV was set to match the multistability experiments in Fig.~\ref{fig3}. Taking the cross-sectional area $A$ of each cavity into account, the two-dimensional polariton-polariton interaction constant is $0.8$ $\mu$eV $\cdot$ $\mu$m$^2$. Dividing by $|X|^2 =  0.16^2$ we get 30 $\mu$eV $\cdot$  $\mu$m$^2$ for the pure exciton-exciton interaction constant. A similar value for the  exciton-exciton interaction constant has been  theoretically estimated in Ref.~\onlinecite{Ciuti98}. \\

%

\vspace{5mm}
\noindent {\large \textbf{Acknowledgements}}

\noindent We thank I. Carusotto, W. Casteels, and C. Ciuti for fruitful discussions, and F. Marsault and F. Baboux, for helpful technical assistance. This work was supported by the
Marie Curie individual fellowship PINQUAR, a public grant over-seen by the French National Research Agency (ANR) as part of the ``Investissements d'Avenir'' program (LabexNanoSaclay, reference: ANR-10-LABX-0035), the ANR project Quandyde (Grant No. ANR-11-BS10-001), the French RENATECH network, and the European Research Council grant Honeypol.\\

\noindent {\large  \textbf{Author contributions}}

\noindent S.R.K.R. performed the experiments, calculations, analyzed the data, and wrote the manuscript. S.R.K.R., A.A., and J.B. designed the experiment, discussed the physics and the manuscript. I.S., L.L., E.G., and A.L. fabricated the sample. \\

\noindent {\large  \textbf{Additional Information}}
Supplementary information is available in the online version of the paper.  Correspondence should be addressed to said.rodriguez@lpn.cnrs.fr\\

\noindent {\large \textbf{Competing financial interests}}

\noindent The authors declare no competing financial interests.

\clearpage

\renewcommand\thefigure{S\arabic{figure}}
\setcounter{figure}{0}

\section{Supplemental Material}
\subsection{ Exciton polaritons in the 2D cavity}
The microstructure investigated in the main text is fabricated by etching a two-dimensional (2D) cavity into the shape of two coupled cavities [Fig. 1(a) in the main text]. Some parts of the 2D cavity are etched on a much larger scale (e.g. 200 microns), leaving behind structures which display the same optical properties as the unetched cavity. Here we analyze the exciton polariton dispersion in such an effectively 2D cavity, where the lateral confinement energy is negligible. Through this analysis, we estimate the exciton and photon fractions of polaritons in the microstructure.

\begin{figure}[H] \centerline{{\includegraphics[width=1\linewidth]{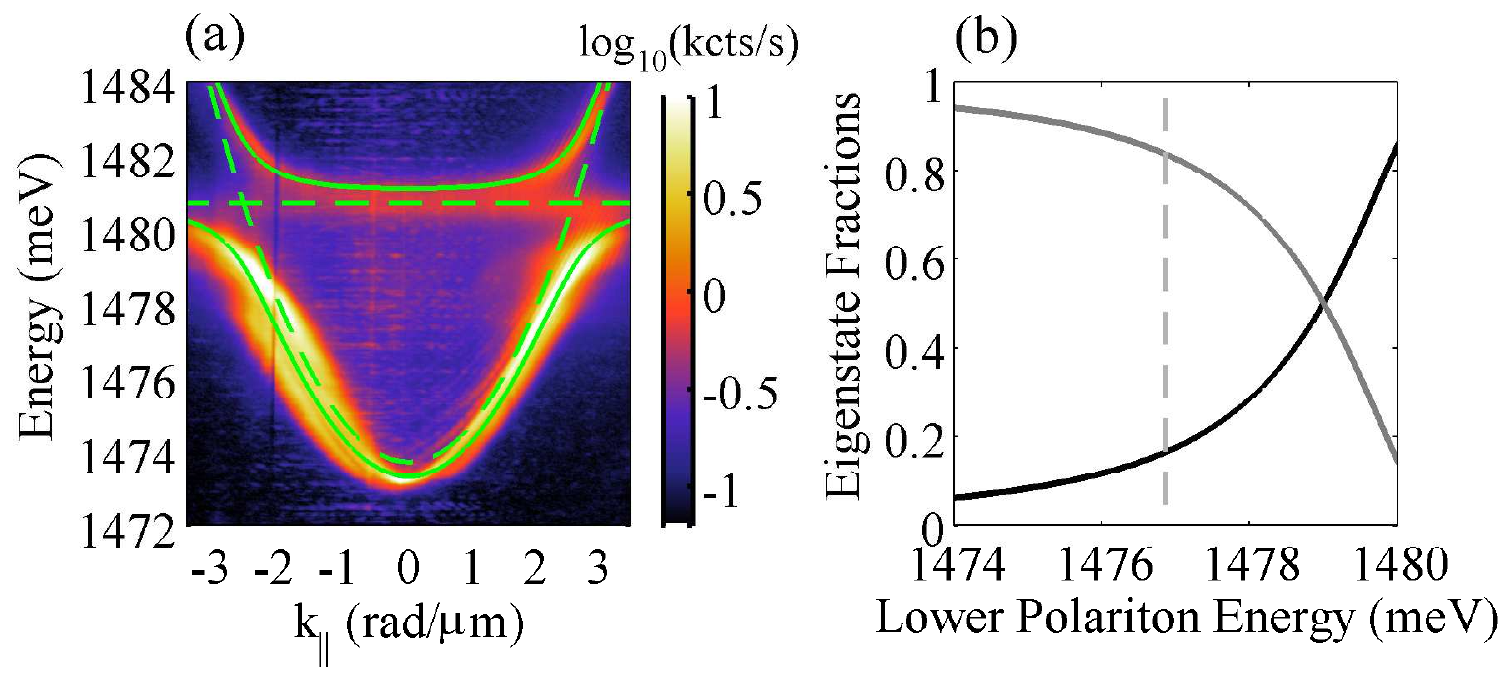}}}\caption{\textbf{Polariton dispersion and eigenstate fractions.} (a) Non-resonant photoluminescence measurements of an effectively two-dimensional  cavity.  The horizontal and parabolic dashed lines indicate the bare exciton and photon energies, respectively. The  solid lines are the upper and lower polaritons formed by strong exciton-photon coupling. (b) For the lower polariton, the exciton fraction is shown as a black line and the photon fraction is shown as a gray line, both as a function of the lower polariton energy. }\label{figS0}
\end{figure}

Figure~\ref{figS0}(a) shows non-resonant photoluminescence measurements of the aforementioned effectively 2D cavity.  The cavity is pumped by a continuous wave laser with an energy of 1.61 eV. The emitted intensity is analyzed spectrally and angularly by imaging the back focal plane of the objective onto a CCD camera.  Figure~\ref{figS0}(a) shows the emitted intensity (in log scale) in color as a function of the emission energy and wave vector component parallel to the quantum well plane. The measurements show the avoided resonance crossing characteristic of exciton polaritons, formed by the strong coupling between cavity photons and quantum well excitons.

We model the polariton system with the following effective $2\times2$ Hamiltonian,

\begin{equation}\label{H}
H = \left( \begin{matrix}
E_{X}  &     \Omega   \\ \nonumber
 \Omega            &  E_{C}(k_\|)    \end{matrix} \right).
\end{equation}

\noindent  $E_X$ and $E_C$($k_\|$)  are the bare exciton and photon energy, respectively; these are the horizontal and parabolic dashed lines Figure~\ref{figS0}(a).  The off-diagonal terms of the Hamiltonian quantify the strength of the exciton-photon coupling --- a fit parameter here set to $\Omega = 1.7$ meV. We calculate the polariton dispersion, i.e. the energy of the mixed states as a function of $k_\|$, by diagonalizing the Hamiltonian. The solid lines in Figure~\ref{figS0}(a) indicate the calculated upper and lower polariton dispersion. The polariton eigenstates can be expressed as $ \left | \wp  \right \rangle = X(k_\|) \left | X \right \rangle +  C(k_\|) \left | C \right \rangle$, with $\left | X \right \rangle$ and $\left | C \right \rangle$ the exciton and cavity photon states. $X(k_\|)$ and $C(k_\|)$ are the Hopfield coefficients  of the eigenvector associated with the eigenvalues of the Hamiltonian (the polariton energy). The magnitude squared of these coefficients, i.e. $|X(k_\|)|^2$ and $|C(k_\|)|^2$, are the eigenstate fractions characterizing the polariton admixture.  In Figure~\ref{figS0}(b) we plot the exciton fraction as a black line and the photon fraction as a gray line, both for the lower polariton as a function of its energy. The vertical dashed line in Fig.~\ref{figS0}(b) indicates the driving energy of the experiments in Fig. 3 and Fig. 4 in the main text. At this energy, the lower polariton has a photon fraction of $|C(k_\|)|^2 = 0.84$.

\subsection{Monostable regime}

\begin{figure}[H] \centerline{{\includegraphics[width=1\linewidth]{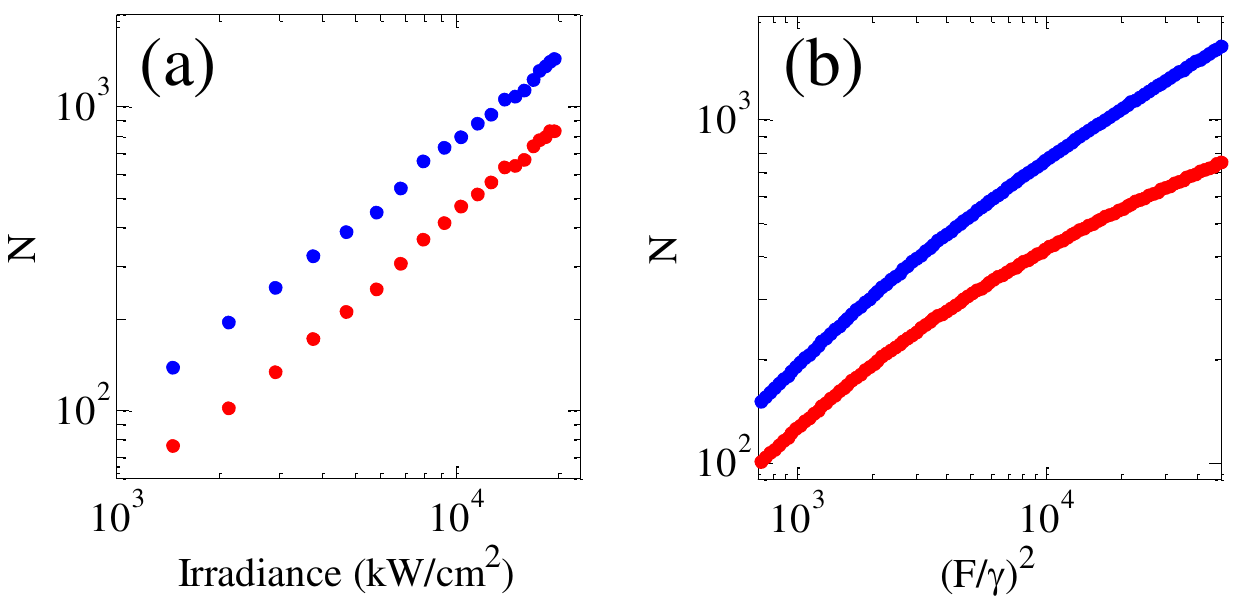}}}\caption{\textbf{Monostable regime.} (a) measurements, and (b) calculations of the population $N$ in each cavity at a driving energy of 1476.36 meV. The blue color corresponds to the driven cavity, and the red color  correspond to the undriven cavity. For the calculations, $F$ is the driving amplitude and $\gamma$ is the bare cavity linewidth.}\label{figS1}
\end{figure}

For a driving energy below the bonding mode energy, each cavity comprising the photonic molecule (PM) exhibits a single input-output branch which is stable for all driving strengths.  In Fig.~\ref{figS1} we show experiments and calculations when driving the left cavity at an energy of 1476.36, red-detuned from the peak bonding energy. Both measurements and calculations show that there is a single input-output branch for all powers. In addition, for very strong driving the calculation shows that the population in the undriven cavity saturates. This power  limiting effect is due to the fact that all modes lie at energies above the driving energy, and that they nonlinearly blue-shift for increasing driving strength due to repulsive interactions. Thus,  this configuration is often called the optical limiter~\cite{Sarchi08}.

\subsection{Bistable regime}
\begin{figure}[H] \centerline{{\includegraphics[width=0.999\linewidth]{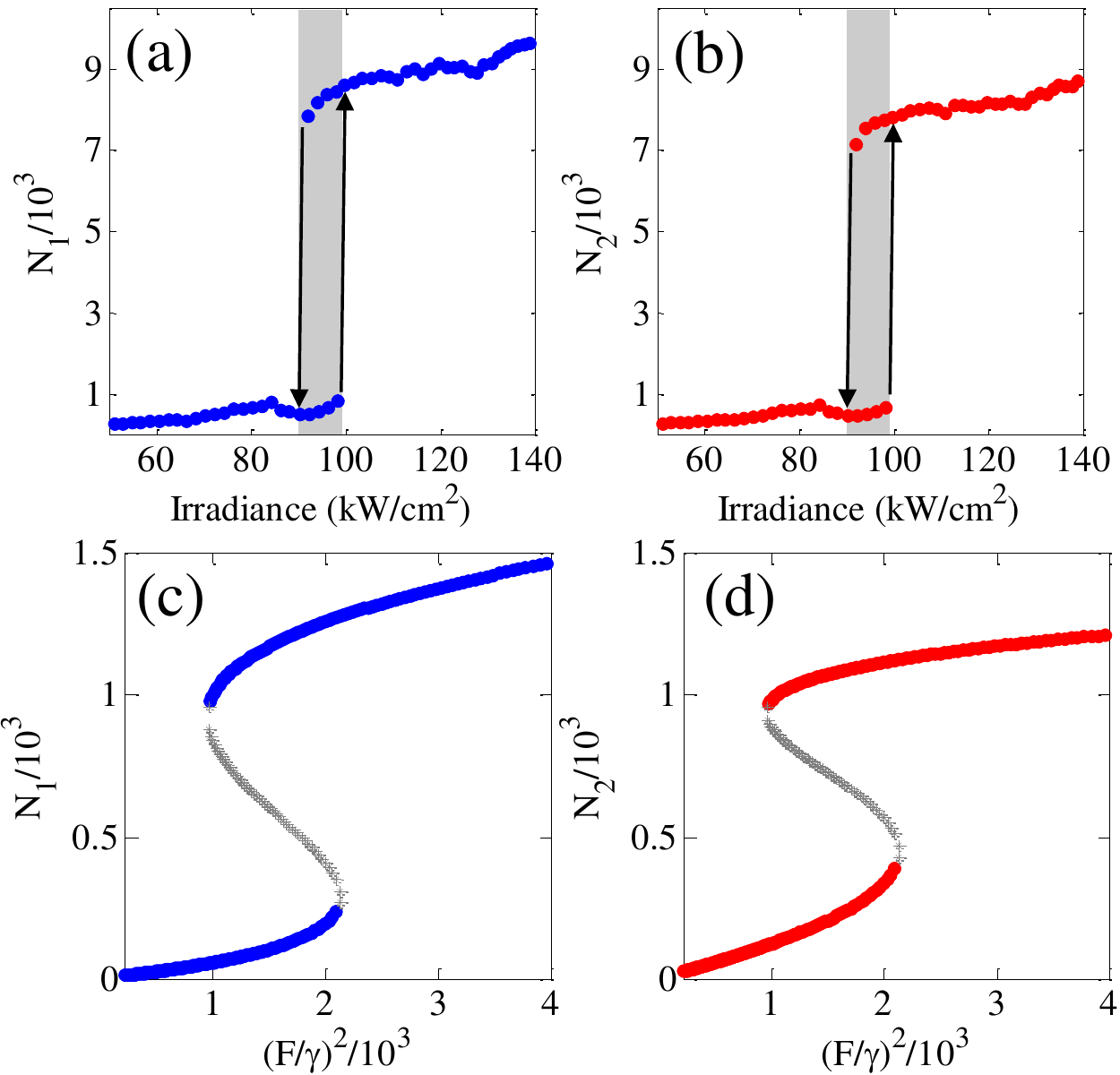}}}\caption{\textbf{Bistability.} (a) and (b) show the measured populations in the driven and undriven cavities respectively, when driving the photonic molecule at 1476.46 meV; (c) and (d)  show the corresponding calculated populations using the model described in the text.  Blue and red data points are stable solutions while gray ones are unstable solutions.}\label{figS2}
\end{figure}

For a driving energy between the  bonding and antibonding mode energies, the PM exhibits up to two stable branches at the same irradiance, i.e. bistability. Figures~\ref{figS2}(a,b) show an example of bistability in the intensity-dependent population of each cavity when driving the left cavity with an  energy of 1476.46 meV.  The shaded ares indicate the bistable region. The calculations obtained using the model equations~\ref{eq1} and shown in Figs.~\ref{figS2}(c,d) give good overall description of this bistable regime.\\

\subsection{Calculated spectrum}

\begin{figure}[H] \centerline{{\includegraphics[width=1\linewidth]{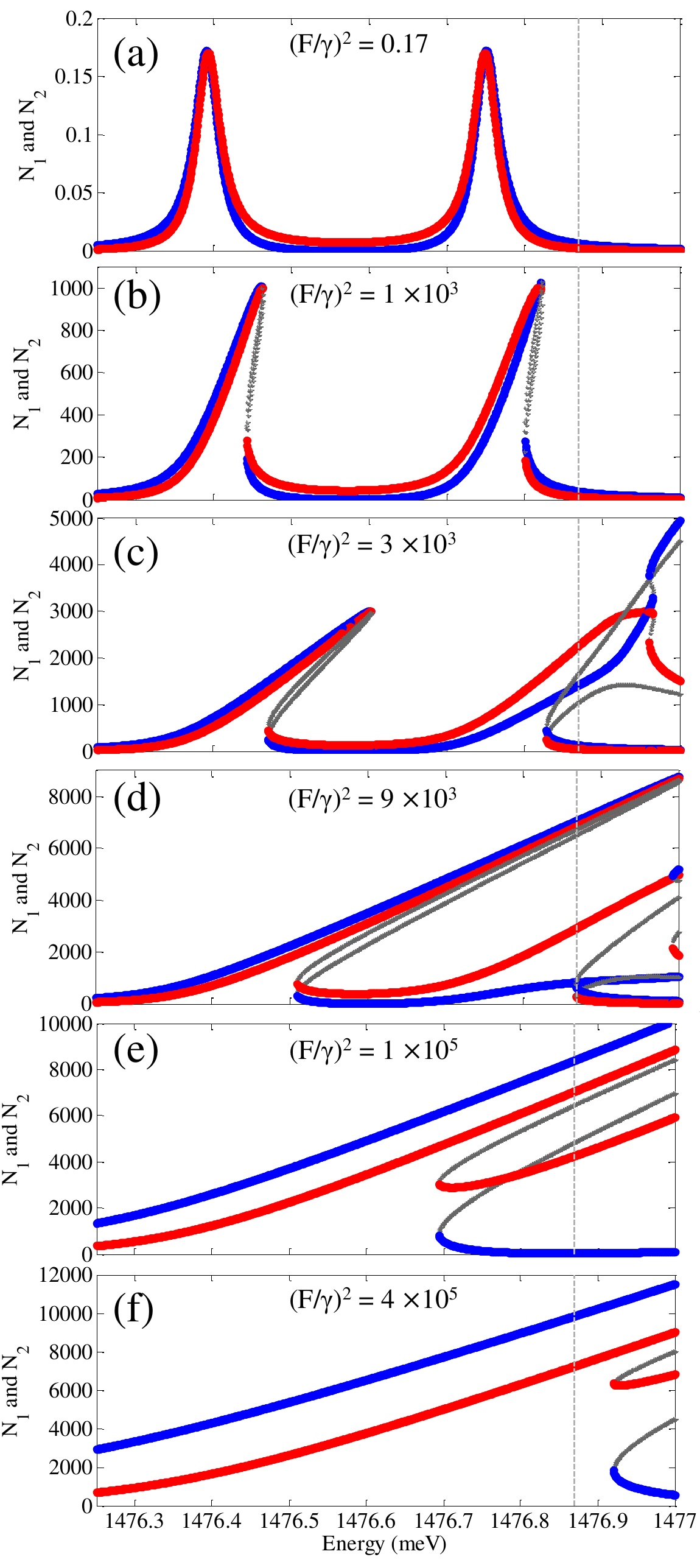}}}\caption{\textbf{Cavity-resolved power-dependent spectrum.} Calculated population of the driven cavity $N_1$ (blue lines) and of the undriven cavity $N_2$ (red lines) for a normalized driving power $(F/\gamma)^2$  indicated at the top of each panel. $F$ is the driving amplitude and $\gamma$ is the bare cavity linewidth. Gray lines correspond to unstable solutions.  The vertical line in all panels indicates the driving energy used for the experiments in Fig. 3 and Fig. 4 of the main manuscript.  }\label{figS4}
\end{figure}

Figure~\ref{figS4} shows the calculated spectrum of the PM for various driving powers. Blue lines indicate the population in the driven cavity, red lines indicate the population in the undriven cavity,  and gray lines indicate unstable solutions. Figure~\ref{figS4}(a) shows the spectrum of the PM in the linear regime (weak driving), here obtained for $(F/\gamma)^2 = 0.17$. The spectrum in Fig.~\ref{figS4}(a) is in good agreement with the experimental result in Fig. 1(a) of the main manuscript. This agreement validates the model parameters retrieved from the Lorentzian fits to the bonding and antibonding resonances in Fig. 1(a). In Fig.~\ref{figS4}(b) we calculate the spectrum for $(F/\gamma)^2 = 10^3$, where single mode bistability is observed.  This corresponds to the regime studied in section C of the supplemental information.   In Fig.~\ref{figS4}(c) obtained for $(F/\gamma)^2 = 3\times 10^3$,  tristability appears  at high energies. For the driving energy of the experiments in Fig. 3 and Fig. 4 in the main manuscript [indicated by the dashed line in all panels of Fig.~\ref{figS4}], there are two stable modes. In Fig.~\ref{figS4}(d) obtained for $(F/\gamma)^2 = 9\times 10^3$, the PM is tristable at the driving energy of the experiments  in Fig. 3 and Fig. 4. In Fig.~\ref{figS4}(e) obtained for $(F/\gamma)^2 = 10^5$, the PM becomes bistable again at the same energy.  The driving power in Fig.~\ref{figS4}(e) corresponds to the dip in population of the driven cavity due to destructive interference. Notice how the lower branch of the driven cavity lies very close to zero population for a wide range of driving energies. This result suggests that the interaction-induced destructive interference effect described in the main manuscript is  robust to relatively large changes in energy (up to several linewidths). Finally, Fig.~\ref{figS4}(f) obtained for $(F/\gamma)^2 = 4 \times 10^5$ shows that for very strong driving the PM becomes monostable at the energy of the dashed line. Thus, the series of calculations in Figure~\ref{figS4} illustrates how the spectrum of the PM evolves as the driving power and interaction energy increase, and how the number of stable modes changes.

\end{document}